\documentstyle[preprint,aps]{revtex}

\begin{document}
\draft
\title{Asymptotic formulae for the Lyapunov spectrum of 
fully-developed shell model turbulence}
\author{Michio Yamada} 
\address{Graduate School of
Mathematical Sciences, University of Tokyo, Tokyo 153,
Japan} 
\author{Koji Ohkitani} 
\address{Faculty of Integrated Arts and Sciences, Hiroshima
University, Higashi-Hiroshima 739, Japan}
\date{$\;\;\;\;\;\;\;\;\;\;\;\;\;\;\;\;\;\;\;\;\;\;\;\;\;\;\;\;\;\;$}
\maketitle

\begin{abstract}

We study the scaling behavior of the Lyapunov spectra of a
chaotic shell model for 3D turbulence.  First, we quantify
localization of the Lyapunov vectors in the wavenumber space
by using the numerical results. Using dimensional arguments
of Kolmogorov-type, we then deduce explicitly the asymptotic
scaling behavior of the Lyapunov spectra. This in turn is
confirmed by numerical results.  This shell model may be
regarded as a rare example of high-dimensional chaotic
systems for which an analytic expression is known for the
Lyapunov spectrum. Implications for the Navier-Stokes
turbulence is given. In particular we conjecture that the
distribution of Lyapunov exponents is {\it not} singular at
null exponent.

\end{abstract}
\pacs{47.27.Jv, 05.45.+b}


We consider the behavior of Lyapunov spectra of a Gledzer's
type shell model of 3D turbulence [1]. In this chaotic
model, a set of complex variables $u_j, \;\; (j=1,2,...,N)$,
which represents the velocity in the shell $k_j=2^{j-10}$,
are governed by the following equations of motion:

$$\left({d \over dt} + \nu k_j^2 \right) u_j ={\rm i}[a_j
u_{j+1}u_{j+2}+b_j u_{j-1}u_{j+1} +c_j u_{j-1} u_{j-2}]^* +
f \delta_{j,4}.\eqno(1)$$ 

The coupling constants in the nonlinear terms are assumed as
follows to ensure energy conservation: $a_j=k_j,\;b_j=-{1
\over 2}k_{j-1},\;c_j={1 \over 2}k_{j-2}$
$b_1=c_1=c_2=a_{N-1}=a_N=b_N=0$.

A number of remarkable properties were revealed by extensive
numerical and analytical studies on this model. In
particular, for large Reynolds $R(=1 / \nu)$ number the
solution to this model is generally chaotic [2-3] and that
its energy spectrum satisfies a Kolmogorov's scaling law of
realistic turbulence.  Other aspects of this model such as
intermittency, probability distributions of velocity
variables [4-6] and the effects of extra helicity-like
invariant are also discussed [7,8].

Because these properties can be achieved with 30-50 degrees
of freedom, its detailed study is feasible to explore
possible links between the conventional theory of turbulence
and the chaotic dynamical systems. One of the most
remarkable properties is that the distribution function of
Lyapunov exponents $\lambda$ appears to diverge at
$\lambda=0$ in the limit of large Reynolds number. This
possibility was pointed out before in [9] using the $\beta -
$ model.  This suggests that the inertial subrange is
connected with a large number of (almost) null exponents.
Indeed, a correlation between Fourier and Lyapunov indexes
was observed in the long-time average of the (squared)
Lyapunov vectors.  However, neither its relation to the
Kolmogorov scaling is left unexplained nor the mechanism of
accumulation of null exponents.  In this Letter, we will see
that how this characteristic Lyapunov spectrum is related
with Kolmogorov's scaling through localization of Lyapunov
vectors in the wavenumber space and obtain an asymptotic
formula for the Lyapunov exponents.

The equations (1) are integrated numerically by the
fourth-order Runge-Kutta method, together with $2N$
linearized equations. Numerical parameters $\nu=10^{-9},
10^{-8}, 10^{-7}, 10^{-6}$ respectively for $N=19, 22, 24,
27$. The forcing is fixed as $f=5\times10^{-3}\times(1 +
i)$.  The time step used was $\Delta t=5 \times 10^{-5}$ for
$N=27$.  After some transient stage, the solution apparently
reaches a statistically stationary state.  Below we will
consider the long-time averages over this state. The energy
spectrum shows about 2 decades of Kolmogorov range (not
shown).  The Lyapunov dimensions are $D=19.8476, 25.4072,
30.2088, 34.9668$ respectively for $N=19, 22, 24, 27$.

Let $v^{(j)}_n$ be the $j-$th Lyapunov vector for the $n-$th
Fourier mode $(j,n= 1, 2,...N)$.  We plot in Fig.1 the
squared components of the Lyapunov vectors in time average
for $N=27$:
$$E^{(j)}(k_n)=< |v^{(j)}_n|^2>.\eqno(2)$$
Note that each Lyapunov vector is normalized as $\sum_{n,j}
|v^{(j)}_n|^2 =1.$ Several distinct features are noted.

\begin{enumerate} 
\item Each Lyapunov vector has a support localized around a 
specific wavenumber.

\item The center of the support of the Lyapunov vector lies
at $ n \approx D/2$ for the largest Lyapunov exponent
($j=1$).

\item The central wavenumber of the support decreases with
$j$, reaching $n \approx 0$. The corresponding Lyapunov
exponents are positive but those corresponding to $\approx
0$ are small.

\item For larger $j$, the central wavenumber increases again
reaching $n \approx D/2$ at $ j \approx D$. The Lyapunov
exponents for these are negative.

\item For even larger $j$, the central wavenumber increases
beyond $ n \geq D/2$. These Lyapunov exponents
asymptotically agree with the reciprocal of the viscous time
scale $- \nu k_n^2$ of the equation (1).

\end{enumerate} 

To summerize, for $0 \leq n \leq D/2$, there are two
Lyapunov vectors for each $n$, one corresponding to positive
exponent and the other negative.  All these features are
consistent with the fact that the Lyapunov dimension
measures the number of modes below the dissipation
wavenumber and that each wavenumber has 2 degrees of
freedom, the real and the imaginary parts of the velocity
variables.

On the basis of the above observation we will introduce the
following set of hypotheses regarding the Lyapunov vectors
in the inertial subrange $j \leq D$ with $D \gg 1$.
\begin{enumerate} 
\item Each Lyapunov vector  in wavenumber space is supported in a 
localized manner around a specific wavenumber.

\item Lyapunov exponents  are positive for $1 \leq j \leq D/2$ and
negative for $D/2 < j$ .

\item Let $n_j$ be the wavenumber for $j-$th Lyapunov vector, then
\begin{enumerate} 
\item $n_j=D/2-j+1$ for $1 \leq j \leq D/2,$
\item $n_j=j-D/2$ for $D/2 \leq j \leq D.$
\end{enumerate} 

\item In the inertial subrange, the $j-$th Lyapunov exponent
($j \leq D$) is inversely proportional to the time scale
$\epsilon^{-1/3} k_{n_j}^{-2/3},$ which is characteristic to
wavenumber $k_{n_j}=k_0 2^{n_j}$.
\end{enumerate} 

The last hypothesis is equivalent to assume that, through
the localization of the Lyapunov vectors in wavenumber
space, the Lyapunov exponents can be expressed solely in
terms of the energy dissipation rate $\epsilon$ and
wavenumber $k$ by the dimensional arguments of
Kolmogorov-type. Combining these hypotheses, we deduce the
following formulae for the Lyapunov exponents $\lambda_j$:

$$\lambda_j \sim  \left\{
\begin{array}{ll}
\epsilon^{1/3} k_{n_j}^{2/3} \sim 2^{-2j/3}, &\quad\mbox{for
$1 \leq j \leq D/2$}\\ -\epsilon^{1/3} k_{n_j}^{2/3}\sim -
2^{2(j-D/2+1)/3}&\quad \mbox{for $D/2 \leq j
\stackrel{<}{\sim} D$}
\end{array}\right. \eqno(3)$$
 
We rescale the Lyapunov exponent as $\lambda_j/H$, where
$H=\sum_{\lambda_j >0} \lambda_j$ is the Kolmogorov-Sinai
entropy. This is equivalent to choose time $\tau=a t \;
(a=\sum_{j=1}^{D/2} \lambda_j)$, such that $H$ is
normalized.  Note that this choice of time neither
influences geometric structure of strange attractors nor
invariant measures on them. Noting that
$$H=\sum_{j=1}^{D/2} \lambda_j \sim { {2^{D/3}-1} \over
2^{2/3} -1 }\eqno(4)$$
we conclude that 
$$\lambda_j / H \sim  \left\{
\begin{array}{ll}
\displaystyle{(2^{2/3}-1) {2^{D/3} \over {2^{D/3}-1}}
2^{-2j/3}} \;\; (D \rightarrow \infty), &\quad\mbox{for $1
\leq j \leq D/2$}\\ \noalign{\vskip 0.2cm}
\displaystyle{2^{2(j-D/2+1)/3} (2^{2/3}-1) \over
2^{2/3}(2^{D/3}-1)} &\quad \mbox{for $D/2 \leq j
\stackrel{<}{\sim} D$}
\end{array}\right. \eqno(5)$$
Because these expressions are free from arbitrary
parameters, we can compare this phenomenological argument
against the numerical results.

The Lyapunov spectra were computed for four values of
viscosity.  In Fig.2 we plot the positive Lyapunov exponents
$\lambda_j / H$ against $H$ together with the theoretical
curve for $D \ll 1$:
$$\lambda_j /H \sim (2^{2/3}-1)2^{-2j/3},\eqno(6)$$ which
can be obtained from (5).  Note that better agreement is
obtained between the phenomenological theory and numerical
results for larger $N$, that is for larger Lyapunov
dimensions.  In Fig.4, we plot $\sum_{i=1}^j \lambda_i / H$
and compare with the theoretical prediction
$$\sum_{i=1}^j \lambda_i / H= 1-2^{-2j/3}. \eqno(7)$$ Again,
the theoretical prediction agree well with the numerical
data and this agreement is better for larger $N$.  Note that
this scaling is different from the one proposed in [2]
previously.  In Fig.4 we replotted the cumulated Lyapunov
spectra using older scaling: $\sum_{i=1}^j \lambda_i / H$ vs
$j/D$.  We see sizable scatters in this enlarged figure and
conclude that the older scaling is premature.

For the other hand for $j > D/2$, a rough the agreement is
seen between the theory and the numerical results. At
present we do not know why the agreement is less clear.

We discuss the implications of the above results for the
realistic Navier-Stokes turbulence.  Suppose that the
support of the Lyapunov vectors are localized in the
wavenumber space in the Navier-Stokes turbulence, as was
true for the case of the shell model.  This implies the
existence of characteristic time scale in the motion
represented by the Lyapunov vectors and that Lyapunov
exponents are given by the reciprocal of relevant time
scales. For simplicity, we consider Navier-Stokes flow under
periodic boundary condition in a box of size $L^3$. The
Reynolds number is assumed to be so large that the motion we
consider is in the inertial subrange. The total number of
eddies of size $r=L/n$ is estimated as $n^3$, where $n$ is a
natural number. This is proportional to the number of modes
whose scale is larger than $r$. The Lyapunov exponent
associated with the smallest eddies can be estimated as the
reciprocal of the Kolmogorov time-scale, that is, $\lambda
\sim r^{-2/3} \sim n^{2/3}.$ In this way we find
$$n^3 \sim \lambda^{9/2}.\eqno(8)$$ This shows that the
number of modes whose Lyapunov exponents do not exceed
$\lambda$ is $\lambda^{9/2}.$ Therefore, the distribution
function $P(\lambda)$ of the Lyapunov exponents is
$$P(\lambda)  \sim \lambda^{7/2} \eqno(9)$$

It should be noted that this distribution $P(\lambda)$ does
not diverge at $\lambda=0,$ in a marked contrast to the case
of shell model. In the latter $P(\lambda)$ has a singularity
at $\lambda=0$ like $1/\lambda.$ In retrospect, The
divergence of the distribution function of Lyapunov
exponents stems from the condensation of modes at null
wavenumber, which is a result of octave discretization of
wavenumbers.  Note that Ruelle's argument uses the
$\beta-$model, whose wavenumbers are also discretized in
octaves. For realistic fluid turbulence, we conjecture that
{\it the distribution function of Lyapunov exponents is
regular at $\lambda=0$} because of no condensation at null
wavenumber.  Unfortunately, it is practically impossible to
check the prediction (9) by the direct numerical
simulations. Further studies on the accumulation of null
Lyapunov exponents, e.g. in connection with the Anderson
localization, will be reported elsewhere.


%
%
\begin{figure}
\caption{Time average of squared components of Lyapunov
vectors : $< |v^{(j)}_n|^2>$.  Contour levels are $0.0489
\times i\;i=1,2,...,10$ Two straight lines represent
correspondence assumed in the hypotheses 3.(a) and (b).}
\label{1}
\end{figure}

\begin{figure}
\caption{Distribution of Lyapunov exponents $\lambda_j/ H$.
double circles ($N=19$), double squares($N=22$), solid
squares ($N=24$) and solid circles ($N=27$). The dashed line
denotes the theoretical prediction : $(2^{2/3}-1)2^{-2j/3}$.
}
\label{2}
\end{figure}

\begin{figure}
\caption{Cumulated distribution of Lyapunov exponents
$\sum_i^j \lambda_i/ H$ Symbols are the same as in Fig.2.
The dashed line denotes the theoretical prediction :
$1-2^{-2j/3}$.}
\label{3}
\end{figure}

\begin{figure}
\caption{Cumulated distribution of Lyapunov exponents
 depicted using previous scaling: $\sum_i^j \lambda_i/ H$
 against $j/D$. Symbols are the same as in Fig.2.}
\label{4}
\end{figure}

\begin{figure}
\caption{Distribution of negative Lyapunov exponents
$|\lambda_j| / H$ against $j-D/2$.  The dashed line denotes
the theoretical prediction : $(2^{2/3}-1)/(2^{2/3}
(2^{D/3}-1))2^{2(j-D/2)/3}$.}
\label{5}
\end{figure}

\begin{references}

\bibitem{1} E. B. Gledzer, Sov. Phys. Dokl.{\bf 18}, 216(1973).

\bibitem{2} M. Yamada  and K. Ohkitani, J. Phys. Soc. Jpn., {\bf 56},
4210(1987).

\bibitem{3} K. Ohkitani and M. Yamada, Prog. Theor. Phys., {\bf 81},
329(1989).

\bibitem{4} M.H. Jensen, D. Paladin and A. Vulpiani,
Phys. Rev.A, {\bf 43}, 798(1991).

\bibitem{5} D. Pisarenko, L. Biferale, D. Courvoisier,
U. Frisch and M. Vergassola, Phys. Fluids A{\bf
5}(1993)2533.

\bibitem{6}  R. Benzi, L. Biferale and G. Parisi,
'On intermittency in a cascade model for turbulence,'
Physica D (Amsterdam) 65(1993)163

\bibitem{7}  L. Kadanoff, D. Lohse, J. Wang, and R. Benzi 
 Phys. Fluids {\bf 7}, 617(1995).

\bibitem{8}  L. Biferale, A. Lambert, R. Lima and G. Paladin,
Physica (Amsterdam) D {\bf 80}105(1995).

\bibitem{9} D. Ruelle,  Commun. Math. Phys. {\bf 87},
287(1982).

\end{references}
\end{document}